\documentclass[prd,amsmath,
twocolumn,floatfix,amssymb, preprintnumbers, nofootinbib, superscriptaddress]{revtex4} 

\usepackage{epsfig}
\usepackage{amsmath}
\usepackage{calrsfs}
\usepackage{nicefrac}
\usepackage{esint}
\usepackage{color}
\usepackage{comment}

\def\barr{\left(\begin{array}{c}}
\def\earr{\end{array}\right)}
\def\bmat{\left(\begin{array}{cc}}
\def\emat{\end{array}\right)}

\def\beq{\begin{equation}}
\def\eeq{\end{equation}}
\def\bea{\begin{eqnarray}}
\def\eea{\end{eqnarray}}
\def\beqa{\begin{equation}\begin{array}{l}}
\def\eeqa{\end{array}\end{equation}}

\begin{document}

\title{Light-by-light forward scattering sum rules for charmonium states}

\author{Igor Danilkin}\email{danilkin@uni-mainz.de}
\author{Marc Vanderhaeghen}
\affiliation{Institut f\"ur Kernphysik \& PRISMA  Cluster of Excellence, Johannes Gutenberg Universit\"at,  D-55099 Mainz, Germany}

\date{\today}

\begin{abstract}
We apply three forward light-by-light scattering sum rules to charmonium states. We show that these sum rules imply a cancellation between charmonium bound state contributions, which are mostly known from the $\gamma \gamma$ decay widths of these states, and continuum contributions above $D \bar D$ threshold, for which we provide a duality estimate. We also show that two of these sum rules allow to predict the yet unmeasured $\gamma^\ast \gamma$ coupling of the $\chi_{c1}(1P)$ state, which can be tested at present high-luminosity $e^+ e^-$ colliders.
\end{abstract}

\maketitle

\section{Introduction}

In recent years, experiments at high luminosity $e^+ e^-$ colliders such as BaBar, Belle, and BESIII have provided a wealth of new meson structure data using the $\gamma \gamma$-fusion process, see e.g. Refs.~\cite{Chernyak:2014wra,UEHARA:2014ura} for some recent reviews. When one or both photons are virtual, such processes allow us to access transition form factors of mesons. Such observables can be interrelated through model independent sum rules for the light-by-light scattering process. In case of the production of hadrons, such relations are shedding light on the non-perturbative dynamics of the underlying theory, Quantum Chromo Dynamics (QCD). 
 
In Ref. \cite{Gerasimov:1973ja,Pascalutsa:2010sj,Pascalutsa:2012pr}, several model independent sum rules for the forward light-by-light scattering were derived and exactly verified at leading order in scalar and spinor QED. Such sum rules are valid for the case when at least one photon is real and the other is spacelike, i.e. for photon virtualities $ q_1^2 = - Q_1^2\leq 0$, $q_2^2 = - Q_2^2 = 0$. Three among these sum rules have the form of a superconvergence relation, for which an integral over an experimentally measurable quantity has to yield zero~\cite{Pascalutsa:2012pr}: 
\begin{eqnarray}
0&=&\int _{s_0}^{\infty }ds\frac{1}{s+Q_1^2}\, 
 \left( \sigma_2 - \sigma_0  \right)_{Q_2^2=0} , \nonumber\\
0&=&\int _{s_0}^{\infty }ds\frac{1}{\left(s+Q_1^2\right)^2}\left(\sigma _{\|}+\sigma _{\text{LT}}+ 
\left(s+Q_1^2\right) \frac{\tau _{\text{TL}}^a}{Q_1Q_2} \right)_{Q_2^2=0} , \nonumber \\
0&=&\int _{s_0}^{\infty }ds\left(\frac{\tau _{\text{TL}}}{Q_1 Q_2}\right)_{Q_2^2=0} , 
\label{eq:sumrules}
\end{eqnarray}
with $\sigma_0$, $\sigma_2$, $\sigma_\|$, $\sigma _{\text{LT}}$, $\tau _{\text{TL}}^a$, and $\tau _{\text{TL}}$ the response functions for the $\gamma^\ast \gamma^\ast \to X$ process, where $X$ denotes the sum over all allowed final states. The experimentally accessible response functions are non-zero above the threshold $s_0$, and are functions of $s = (q_1 + q_2)^2$, $Q_1^2$, and $Q_2^2$, see Ref.~\cite{Budnev:1974de,Pascalutsa:2012pr} for details. In previous works, applications of the light-by-light sum rules to different model field theories have been demonstrated both in perturbative and non-perturbative settings~\cite{Pascalutsa:2010sj,Pascalutsa:2012pr,Pauk:2013hxa}. Furthermore, in Refs.~\cite{Pascalutsa:2012pr,Danilkin:2016hnh}, their application to the $\gamma^\ast \gamma$ production of light-quark mesons have been discussed. For these light pseudo-scalar, scalar, axial-vector, and tensor mesons, where data are available, these sum rules were shown to be verified within the 10 - 30 \% experimental accuracies. Furthermore, an application of the sum rules of Eq.~(\ref{eq:sumrules}) also allowed to predict the $\gamma^\ast \gamma$ transition form factor of the $f_2(1270)$  meson~\cite{Pascalutsa:2012pr}, which was later measured by the Belle Collaboration~\cite{Masuda:2015yoh} and found to be in good agreement with the sum rule prediction~\cite{Danilkin:2016hnh}.  
 
In the present work we will apply the sum rules of Eq.~(\ref{eq:sumrules}) to the charmonium sector. We will show that in order to satisfy these sum rules a cancellation between contributions from charmonium bound states and continuum contributions above $D \bar D$ threshold is required. We will also show that two of these sum rules allow us to make a prediction for the $\gamma^\ast \gamma$ coupling to the axial-vector $\chi_{c1}(1P)$ state, which has not yet been extracted from experiment to date, but is accessible at the present day high-luminosity $e^+e^-$ colliders. Furthermore, the application of light-by-light sum rules in the charmonium sector may be worthwhile in light of the plethora of new states (so-called $XYZ$ states) which have been found in recent years above $D \bar D$ production threshold, of which several have been produced in $\gamma \gamma$ collisions,  see e.g. Refs.~\cite{Olsen:2014qna,Chen:2016qju,Lebed:2016hpi} for some recent reviews.

The outline of this work is as follows. In Section II, we will provide an update of the analysis for the first sum rule in Eq.~(\ref{eq:sumrules}) in the charmonium sector. Subsequently in Section III, we will apply our formalism to the second and third $\gamma \gamma$ sum rules of Eq.~(\ref{eq:sumrules}), which requires one photon to be quasi-real. This will allow us to make a quantitative prediction for the decay width of the  $\chi_{c1}(1P)$ state into a real and a quasi-real photon. We will also be able to provide an error estimate and compare our result with the quark model prediction. Finally we will present our conclusion in Section IV.

 \section{Real photon helicity sum rule for charmonium states}
 
As a first application, the first sum rule of Eq.~(\ref{eq:sumrules}) was tested in \cite{Pascalutsa:2012pr} for the case of real photons 
in the light-quark and charm-quark sectors. In the narrow width approximation it yields a relation between the $\gamma \gamma$ decay widths of pseudoscalar ($\cal{P}$), scalar ($\cal{S}$), and tensor ($\cal{T}$) mesons, denoted by $\Gamma _{\gamma \gamma}({\cal P, S, T})$, as:
\begin{eqnarray}
0&=&\int _{s_0}^{\infty }\frac{ds}{s}(\sigma_2(s)-\sigma_0(s))\nonumber\\
&=&-\sum _{\cal{P}} 16\,\pi ^2\,\frac{\Gamma_{\gamma \gamma} ( \cal{P} ) }{m_{\cal{P}}^3} -\sum _{\cal{S}} 16 \,\pi ^2\,\frac{\Gamma_{\gamma \gamma} ( \cal{S}) }{m_{\cal{S}}^3}
\nonumber\\
&+&\sum _{\cal{T}} 16 \,\pi ^2\,\frac{5\, \Gamma_{\gamma \gamma} ( {\cal{T}} (\Lambda = 2) ) }{m_{\cal{T}}^3}\,,
\end{eqnarray}
with $m_M$ the corresponding meson mass, and where for the tensor mesons we only show the dominant contribution corresponding with helicity $\Lambda = 2$, see Ref.~\cite{Danilkin:2016hnh} for the full expression and extension to virtual photons.

In order to satisfy the helicity sum rule, it was found in \cite{Pascalutsa:2012pr} that there is a quantitative cancellation between $\pi^0$ and $a_2$ in the low-lying isovector sector and a cancellation between $\eta$, $\eta'$ and $f_2(1270)$ in the low-lying isoscalar sector. The charmonium family presents an interesting difference, as the spectrum can be separated into two parts: (narrow) bound state  contributions below $D\bar{D}$ threshold, and resonance and continuum contributions above $D\bar{D}$ threshold. As one can see from Table~\ref{table_ccbar}, from the measured two-photon decay widths, the dominant, i.e. lowest lying, bound state contribution comes from the $\eta_c(1S)$ state, while the lowest lying scalar $\chi_{c0}(1P)$ and tensor $\chi_{c2}(1P)$ charmonia to good approximation cancel each other.

\begin{table*}[t]
{\centering \begin{tabular}{|c|c|c|c|c|}
\hline
& $m_M$   & $\Gamma_{\gamma \gamma} $   
&  \multicolumn{2}{|c|} { $\int \frac{ds}{s}\;  (\sigma_2 - \sigma_0) $ } \\
&  [MeV] &   [keV] 
&  \multicolumn{2}{|c|} {  [nb] } \\
\hline 
\hline
\quad $\eta_c(1S)$ \quad  & \quad  $2983.4 \pm 0.5$ \quad   & \quad  $5.1 \pm 0.4$ \quad   &  
\multicolumn{2}{|c|} { $  -11.8 \pm 0.9$   }\\
\hline
\quad $\chi_{c0}(1P)$  \quad  &  \quad $3414.75 \pm 0.31$  \quad  & \quad  $2.34 \pm 0.19$ \quad &  
\multicolumn{2}{|c|} {  $ -3.6 \pm 0.3 $ } \\
\hline
\quad $\chi_{c2}(1P)$ \quad &  \quad $3556.20 \pm 0.09$  \quad  & \quad  $0.53 \pm 0.04$ \quad   & 
\multicolumn{2}{|c|} {\quad  $3.6 \pm 0.3$    }  \\
\hline
\quad sum of states $1S, 1P$ \quad &  & & \multicolumn{2}{|c|} { $ -11.8 \pm 1.0$ } \\
\hline 
\hline
\quad $\eta_c(2S)$ \quad  & \quad  $3639.2 \pm 1.2$ \quad   & \quad  $1.3 \pm 0.6$ \quad   &  
\multicolumn{2}{|c|} { $ -1.7\pm 0.8 $  } \\
\hline
\quad $\chi_{c0}(2P)$  \quad  &  \quad $3862 \pm 48 $  \quad  & \quad  $  1.2 \pm 0.3$ \quad &  
\multicolumn{2}{|c|} { $ -1.2 \pm 0.3 $ } \\
\hline
\quad $\chi_{c2}(2P)$ \quad &  \quad $3927.2 \pm 2.6$  \quad  & \quad  $ 0.21 \pm 0.04$ \quad   & 
\multicolumn{2}{|c|} {  \quad  $ 1.1\pm 0.2$     } \\
\hline
\quad sum of states $2S, 2P$ \quad &  & & \multicolumn{2}{|c|} { $ -1.8 \pm 0.9$ } \\
\hline
\hline
\quad sum of states $1S, 1P, 2S, 2P$ \quad &  & & \multicolumn{2}{|c|} { $-13.6\pm 1.3$ } \\
\hline
\hline
\quad continuum contribution \quad  &  & &   $\sqrt{s} \geq 2m_D$ &  $\sqrt{s} \geq 4.077$ GeV  \\
 &  & &  $9.0 \pm 0.8$ & $10.0 \pm 0.8$  \\
\hline
 bound states + continuum  &  & &   $-2.8 \pm 1.3$  &  $-3.6\pm 1.5$ \\
\hline
\end{tabular}\par}
\caption{Contributions of the lowest $c \bar c$ mesons to the $\gamma \gamma$ helicity sum rule. We show the bound state contributions  based on the 2016 PDG values~\cite{Olive:2016xmw} of the meson masses $m_M$ and their $2 \gamma$ decay widths $\Gamma_{\gamma \gamma}$. For the $\chi_{c0}(2P)$ and $\chi_{c2}(2P)$ we use estimates from Eqs. (\ref{GammaGammaChic0(2P)}) and (\ref{GammaGammaChic2(2P)}).
Note, that the $\chi_{c0}(2P)$ charmonium state is identified with the recently measured $X^*(3860)$ \cite{Chilikin:2017evr}. For simplicity we symmetrized the error bar of its mass by enlarging the smallest error.
We also show the duality estimate for the continuum contribution, as well as the total sum.}
\label{table_ccbar}
\end{table*}

For the higher states, the two photon decay width is also known for the $\eta_c(2S)$ bound state from the CLEO experiment~\cite{Olive:2016xmw}. For the $\chi_{c0}(2P)$ and $\chi_{c2}(2P)$ states, 
which are lying above $D \bar D$ threshold, 
there are no directly reported values of the two photon decay widths, but they can be approximately estimated. The value $\Gamma_{\gamma\gamma}[\chi_{c2}(2P)]$ can be expressed through the branching fraction for the $\gamma \gamma$ production of this states multiplied by its decay $\chi_{c2}(2P)\to \bar{D}D$. Under the assumption that the latter is the dominant decay mode, we obtain~\cite{Olive:2016xmw}: 
\begin{eqnarray}\label{GammaGammaChic2(2P)}
\Gamma_{\gamma\gamma}[\chi_{c2}(2P)]&\gtrsim& \frac{\Gamma[\chi_{c2}(2P)\to \bar{D}D]}{\Gamma_{tot}}\,\Gamma_{\gamma\gamma}[\chi_{c2}(2P)]\nonumber\\
&=&(0.21\pm 0.04) ~ \text{keV}.
\end{eqnarray}
For the $\chi_{c0}(2P)$ we use several estimates. For a first estimate we identify it with the $X(3915)$ state assuming that its total width is obtained from the dominant $X(3915)\to \omega\, J/\psi$ and $X(3915) \to \bar{D}^{*0}D^0$ decays. Using the measured value~\cite{Olive:2016xmw}: 
\begin{equation}
 \Gamma_{\gamma\gamma}[X(3915)]\,\frac{\Gamma [X(3915)\to \omega\, J/\psi]}{\Gamma_{tot}}=(54\pm 9)~\mathrm{eV},
\end{equation} 
as well as the bound~\cite{Olive:2016xmw}: 
\begin{equation}
 \frac{\Gamma[X(3915)\to \omega\, J/\psi] }{\Gamma[X(3915) \to \bar{D}^{*0}D^0 ]} >0.71,
\end{equation} 
we estimate:
\begin{eqnarray}\label{GammaGammaChic0(2P)_0}
\Gamma_{\gamma\gamma}[X(3915)] &\lesssim & \Gamma_{\gamma\gamma}[X(3915)]\,\frac{\Gamma[X(3915)\to \omega\, J/\psi]}{\Gamma_{tot}}\nonumber \\
& &\times\left(1+ \frac{\Gamma[X(3915) \to \bar{D}^{*0}D^0]}{\Gamma[X(3915)\to \omega\, J/\psi]}\right) \nonumber\\
&\lesssim& (0.13\pm 0.2) ~ \text{keV}. 
\end{eqnarray}
However, as was pointed out in Ref. \cite{Guo:2012tv}, the identification of the $X(3915)$ state with the  $\chi_{c0}(2P)$ charmonium state has several issues: $X(3915)$ was not observed in the $D\bar{D}$ channel, the partial width for the $X(3915)\to \omega J/\psi$ is too large and the mass difference between $X(3915)$ and $\chi_{c2}(2P)$ is too small compared to the expected fine splitting of the 2P levels in charmonium.

We will therefore use a second scenario in the following, in which we identify $\chi_{c0}(2P)$ with the $X^*(3860)$ state. This new charmoniumlike state was observed very recently in the process $e^+e^- \to J/\psi\,D\bar{D}$ by the Belle Collaboration \cite{Chilikin:2017evr}. Its mass was obtained as $3862^{+26+40}_{-32-13}$ MeV and its total width as $201^{+154+88}_{-67-82}$ MeV. As its two photon decay width remains yet to be measured, we will extract it here from the alternative fit to the Belle \cite{Uehara:2005qd} and BABAR \cite{Aubert:2010ab} $\gamma\gamma \to D\bar{D}$ data performed in Ref. \cite{Guo:2012tv}. Assuming that both $\chi_{c0}(2P)$ and $\chi_{c2}(2P)$ states have a dominant decay to $D\bar{D}$, the authors in Ref. \cite{Guo:2012tv} have fitted the $\gamma\gamma \to D\bar{D}$ cross section below $\sqrt{s}\lesssim 4.2$ GeV by a sum of $\chi_{c0}(2P)$ and $\chi_{c2}(2P)$ resonance contributions. By integrating the invariant-mass distribution from the $D\bar{D}$ threshold up to $\sqrt{s_{D_2}}=M_{\chi_{c0}(2P)}+\Gamma_{\chi_{c0}(2P)}=4.077$ GeV one obtains from such fit:
\begin{equation}
\frac{\Gamma_{\gamma\gamma}[\chi_{c0}(2P)]}{\Gamma_{\gamma\gamma}[\chi_{c2}(2P)]}=5.6 \pm 0.9\,.
\end{equation}
By furthermore using our estimate for $\chi_{c2}(2P)$ in Eq. (\ref{GammaGammaChic2(2P)}) we obtain
\begin{equation}\label{GammaGammaChic0(2P)}
\Gamma_{\gamma\gamma}[\chi_{c0}(2P)]=1.2  \pm 0.3~ \text{keV}.
\end{equation}
We show the resulting contributions to the helicity sum rule in Table~\ref{table_ccbar}.

It can be seen that the  that the $\eta_c(2S)$,  $\chi_{c0}(2P)$, and $\chi_{c2}(2P)$ contributions to the helicity sum rule amount to around 15\% of the contributions of the lowest lying $\eta_c(1S)$,  $\chi_{c0}(1P)$, and $\chi_{c2}(1P)$ states. Note that if we assume that the $\chi_{c0}(2P)$ state corresponds to $X(3915)$ then its contribution to the sum rule would be even smaller, $ -0.13 \pm 0.02$.

\begin{figure}[t]
\includegraphics[width =0.48\textwidth]{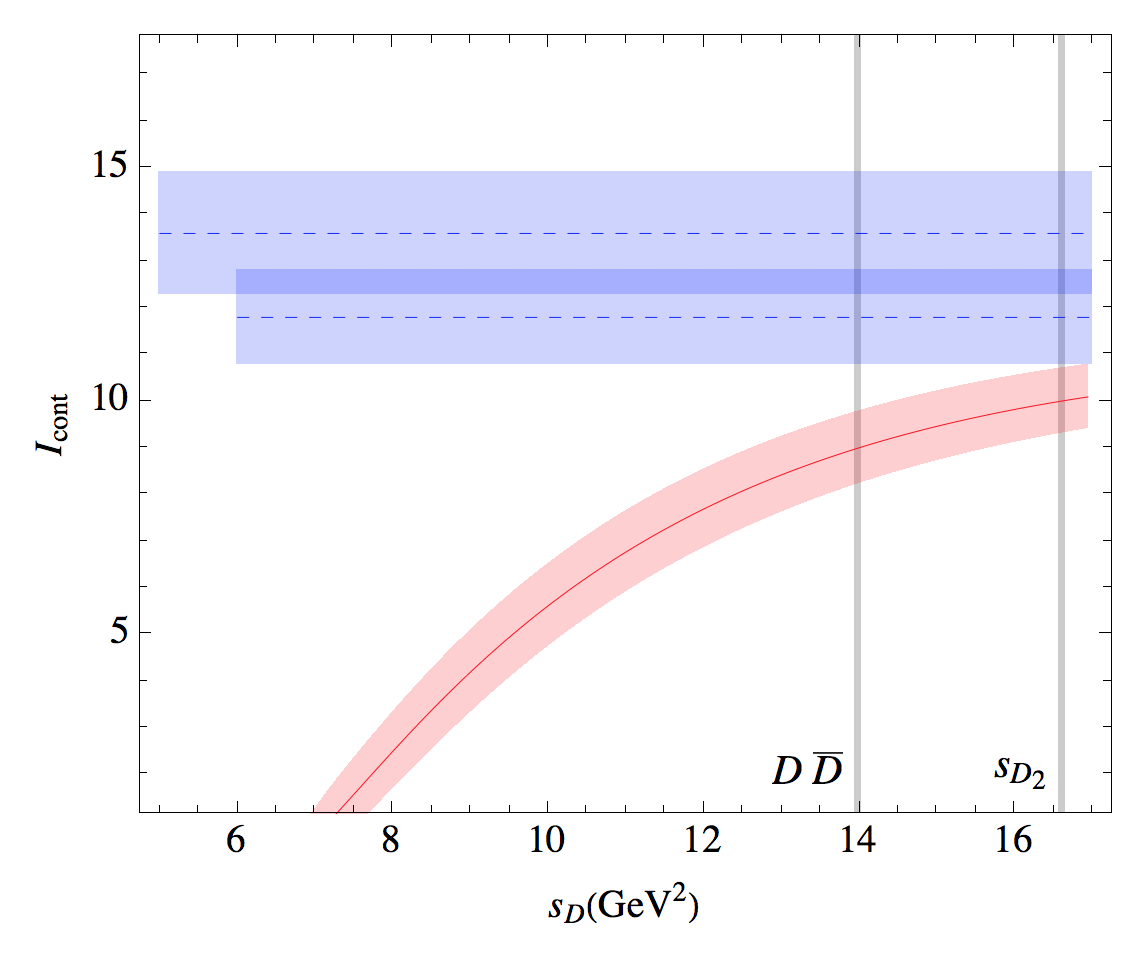}
\caption{Solid (red) band: 
plane wave duality estimate for the continuum contribution $I_{cont} $ of Eq.~(\ref{eq:duality3}) to the helicity difference sum rule for 
charm quarks as function of the open charm production threshold $s_D$. 
The vertical lines indicate the $D \bar D$ threshold, $s_D \equiv 4 m_D^2 \approx 14$~GeV$^2$ and $s_{D_2}=16.6$~GeV$^2$.
The lower (blue) horizontal band indicates the sum of the 
$\eta_c(1S), \chi_{c0}(1P)$, and $\chi_{c2}(1P)$ bound state contributions (taken with the opposite sign) to the $\sigma_2 - \sigma_0$ sum rule, as listed in Table~\ref{table_ccbar}. The slightly higher (blue) horizontal band includes the contributions of the 
 $\eta_c(2S), \chi_{c0}(2P)$, and $\chi_{c2}(2P)$ states. 
The proximity between red and blue bands  is indicative of the compensation between the lowest bound state contributions and the duality estimate for the continuum contribution to the helicity difference sum rule.}
\label{fig:duality}
\end{figure}

In order to estimate the contribution from all states above the open-charm threshold, which opens at $s_D \equiv 4 m_D^2 \approx 14$~GeV$^2$, using the $D$-meson mass $m_D \approx 1.87$~GeV,  a quark-hadron duality argument~\cite{Novikov:1977dq}  was used in \cite{Pascalutsa:2012pr}. 
This duality estimate amounts to replace the helicity-difference $\gamma\gamma\rightarrow X$ cross section, 
entering the continuum integral of the sum rule, by the perturbative $\gamma\gamma\rightarrow Q \bar{Q}$ cross section:
\begin{eqnarray}
I_{cont} &\equiv& \int_{s_D}^{\infty}  \, \frac{d\,s}{s} \left( \sigma_2 - \sigma_0 \right) _{\gamma \gamma \to X}
\nonumber \\
&\approx & \int_{s_D}^{\infty} \, \frac{d\,s}{s} \left( \sigma_2 - \sigma_0 \right)_{\gamma \gamma \to Q\bar Q}
\nonumber\\
&=& 
- \int_{4 m_c^2}^{s_D}  \, \frac{d\,s}{s} \left(\sigma_2 - \sigma_0 \right)_{\gamma \gamma \to Q \bar Q}\,,
\label{eq:duality1}
\end{eqnarray}
with $m_c$ the charm quark mass, and where the last equation follows from the fact that sum rule is satisfied exactly for plane wave states in spinor QED. 
Using the perturbative $\gamma \gamma \to Q \bar Q$ cross section derived in Refs.~\cite{Pascalutsa:2010sj, Pascalutsa:2012pr}, one obtains for the continuum integral~\footnote{Note that in Ref.~\cite{Pascalutsa:2012pr}, the factor $e_c^4 N_c \simeq 0.6$ in the continuum integral was not accounted for. }:
\begin{eqnarray}
I_{cont}
&\approx& - 8 \pi \alpha^2 e_c^4 \, N_c \int \limits_{4 m_c^2}^{s_D}  \frac{ ds }{s^2} 
\Big\{ -3\,\beta(s) \nonumber\\ &&
 \qquad\qquad\qquad+ 2 \ln \left( \frac{\sqrt{s}}{2 m_c}(1+\beta(s))\right)   \Big\},\nonumber\\
\beta(s)&\equiv&\sqrt{1 - \frac{4 m_c^2}{s}}\,,
\label{eq:duality3}
\end{eqnarray}
with $e_c = 2/3$ the charm quark charge, and number of colors $N_c = 3$.
Using the PDG value $m_c = 1.27 \pm 0.03 $~GeV for the $\overline{\rm{MS}}$ charm quark mass \cite{Olive:2016xmw}, Eq.~(\ref{eq:duality3}) yields: 
$I_{cont} \approx 9.0 \pm 0.8$~nb, where the error results from the above uncertainty range in the PDG value for $m_c$. In Fig.~\ref{fig:duality} we also show $I_{cont}$ as a function of the integration limit $s_D$. We see from Table~\ref{table_ccbar} that this continuum estimate for $s_D=4m_D^2$ has the opposite sign from the bound state contributions, and 
compensates to around  75 \% the sum of the $\eta_c(1S), \chi_{c0}(1P)$, and $\chi_{c2}(1P)$ contributions to the $\sigma_2 - \sigma_0$ sum rule. Such cancellation quantitatively illustrates the interplay between charmonium bound states and resonances,  (dominantly) decaying into charmed mesons, in satisfying the sum rules of Eq.~(\ref{eq:sumrules}). 
We furthermore show in Table \ref{table_ccbar} the estimate for the $\eta_c(2S)$, $\chi_{c0}(2P)$, and $\chi_{c2}(2P)$ states. 
In this case, the duality estimate for the continuum states starts above these states in order to avoid double counting, i.e. for $s_{D_2}$. It is remarkable to note that the contribution of these higher states is compensated within the error bar by the duality estimate in the interval $[4m_D^2,\,s_{D_{2}}]$, providing further support of our procedure.
 It will be interesting to test this experimentally by directly measuring the $\gamma \gamma$ production cross sections above $D \bar D$ threshold, where a large number of new states (so-called $XYZ$ states) have been found in recent years, several of which have exotic quantum numbers and are still very poorly understood. 
 
For a more precise estimate of the continuum part of the spectrum one needs to account for the interaction between the quarks. Typically, the heavy quarkonia can be described within a non-relativistic (NR) potential quark model, where the interaction is presented as a sum of the Coulomb potential (one-gluon exchange) and a linear (confinement) term
\begin{equation}
V(r)=-\frac{4}{3}\frac{\alpha_s}{r}+\sigma\,r\,,
\end{equation}
with $\alpha_s$ the strong coupling and $\sigma$ the string tension. One can expect a change in $I_{cont}$ by a Coulombic interaction, since it is supposed to be the dominant interaction for the region $s\in [4m_c^2,\,s_D]$, i.e. below $D\bar{D}$ threshold. 
In the threshold region the Sommerfeld enhancement mechanism~\cite{Sommerfeld} is known to qualitatively change the cross section, and we may therefore expect the continuum contribution to the sum rules to be enhanced as compared with a plane wave cacluation, closing the gap in the sum rule evaluation shown in Table~\ref{table_ccbar}.   
Although in the present work we make the simple duality estimate using plane wave states for the outgoing quarks when evaluating 
the continuum contribution to the sum rules of Eq.~(\ref{eq:sumrules}), 
we plan to perform such a detailed study of the modifications when using continuum Coulombic wave functions in a future work. 

\section{Virtual photon sum rules for charmonium states}

We next discuss the implications of the second and third sum rules of Eq.~(\ref{eq:sumrules}), which we denote by SR$_2$ and SR$_3$ in the following, when both photons are quasi-real for charmonium states. Pseudo-scalar mesons do not contribute to these sum rules, which instead receive contributions from scalar, axial-vector and tensor mesons. To satisfy both of these sum rules implies therefore that there is a  compensation between those meson bound states with the continuum contributions, 
which we will study subsequently. 

Using the cross section expressions~\cite{Pascalutsa:2012pr}, the contributions of narrow scalar ($\cal{S}$), 
axial-vector ($\cal{A}$), and tensor ($\cal{T}$) mesons to SR$_2$ and SR$_3$ for two quasi-real photons were derived in Ref.~\cite{Danilkin:2016hnh}. In applying these results to the charmonium states,  the sum rules SR$_2$ and SR$_3$ can be expressed as a sum over the narrow charmonia bound states and a continuum contribution as:
\begin{eqnarray}\label{sr2meson0}
0&=&\sum _{\cal S} 
16\pi ^2  \frac{\Gamma _{\gamma \gamma }({\cal S})}{m_{\cal S}^5}
\biggl (1-R_{{\cal S} }^L(0) \biggr )  
- \sum _{\cal A}8\pi^2 \frac{3\,\tilde \Gamma_{\gamma\gamma}(\cal A)}{m_{\cal A}^5}
\nonumber\\
&+&\sum_{\cal T} 8\pi^2 \frac{5\,\Gamma _{\gamma \gamma }({\cal T})}{m_{\cal T}^5} 
\Big\{ r^{(2)}  +  r^{(0)} \left( 2 +  R_{{\cal T} }^{L}(0)  \right) 
\\&&\quad 
+ \frac{\pi\alpha^2 m_{\cal T}}{10 \, \Gamma _{\gamma \gamma }({\cal T})} 
\left[ F_{{\cal T} \gamma ^*\gamma ^*}^{(1)}\left(0,0\right)   \right]^2 \Big\} + I_{cont}(\mathrm{SR}_2)\,,\nonumber
\end{eqnarray}
and
\begin{eqnarray}\label{sr3meson0}
0=&-&\sum _{\cal S} 16\pi ^2\,\frac{\Gamma _{\gamma \gamma }({\cal S})}{m_{\cal S}^3} R_{{\cal S} }^L(0)
+ \sum _{\cal A} 8\pi ^2\,\frac{3\,\tilde \Gamma_{\gamma \gamma} ({\cal A})}{m_{\cal A}^3} 
\nonumber\\
&+& \sum_{\cal T} 8\pi^2 \frac{5\,\Gamma _{\gamma \gamma }({\cal T})}{m_{\cal T}^3}
\Big\{ r^{(0)} R_{{\cal T} }^{L}(0) 
\\&&\quad 
- \frac{\pi\alpha ^2\,m_{\cal T}}{10 \, \Gamma _{\gamma \gamma }({\cal T})} 
\left[ F_{{\cal T} \gamma ^*\gamma ^*}^{(1)}\left(0,0\right) \right]^2 
\Big\}\, + I_{cont}(\mathrm{SR}_3)\,,\nonumber
\end{eqnarray}
where $r^{ (\Lambda)}$ denotes the
ratio of the two-photon decay widths of the tensor meson with specific helicity $\Lambda$ relative to the total
two-photon decay width. Furthermore, 
$ R_{{\cal S} }^L(0)$ and $R_{{\cal T} }^L(0) $ denote the longitudinal over transverse $\gamma^\ast \gamma^\ast \to {\cal{S, T}}$ coupling ratios respectively. The equivalent $2 \gamma$ decay width $\tilde \Gamma_{\gamma \gamma} ({\cal A})$ for axial-vector mesons is defined as~\cite{Schuler:1997yw}:  
\begin{eqnarray}
\tilde \Gamma_{ \gamma \gamma}({\cal A}) &\equiv& \lim \limits_{Q_1^2 \to 0} \, \frac{m_{\cal A}^2}{Q_1^2} \, \frac{1}{2} \,
\Gamma \left( {\cal A} \to \gamma^\ast_L \gamma_T \right)\\
&=& \frac{\pi \alpha^2}{4} \, m_{\cal A} \, \frac{1}{3} \left[ F^{(1)}_{{\cal A} \gamma^\ast \gamma^\ast}(0, 0)  \right]^2,\nonumber
\label{a2gwidth}
\end{eqnarray}
with $F^{(1)}_{{\cal A} \gamma^\ast \gamma^\ast}(Q_1^2, Q_2^2)$  the corresponding 
$\gamma^\ast \gamma^\ast \to {\cal{A}}$
transition form factor, see Refs.~\cite{Pascalutsa:2012pr, Danilkin:2016hnh} for details. Furthermore, in Eqs.~(\ref{sr2meson0}) and (\ref{sr3meson0}) the continuum contributions take the form:
\begin{eqnarray}
I_{cont} (\mathrm{SR}_2) &\equiv& \int_{s_D}^{\infty}  \, d\,s \, \left( \frac{\sigma_\parallel}{s^2}   + \frac{1}{s} \frac{\tau^a_{TL}}{Q_1 Q_2} \right)_{Q_i^2  = 0} (\gamma \gamma \to X), \nonumber\\
I_{cont} (\mathrm{SR}_3) &\equiv& \int_{s_D}^{\infty}  \, d\,s \, \left( \frac{\tau_{TL}}{Q_1 Q_2} \right)_{Q_i^2  = 0} (\gamma \gamma \to X)\,. 
\end{eqnarray}

As both sum rules have to integrate to zero, they also  imply a cancellation 
mechanism between the bound state and the continuum contributions. 
We first discuss the contributions of scalar, axial-vector and tensor charmonium bound states  
to SR$_2$ and SR$_3$. 

For the scalar charmonium state $\chi_{c0}(1P)$ the two-photon decay width is known, which determines the transverse coupling. For the L/T ratio 
$ R_{{\cal S} }^L(0) $, we take the quark model prediction, see Eq.(\ref{S:ratios}) in Appendix \ref{Appendix_Quark_Model}, $ R_{{\cal S} }^L(0) = 2/3$. 

For the axial-vector charmonium state $\chi_{c1}(1P)$, for which the equivalent two-photon decay width $\tilde \Gamma_{\gamma \gamma}$ is not known, we will express the sum rule contribution as a function of the branching fraction $\tilde \Gamma_{\gamma \gamma} / \Gamma_{tot}$, with $ \Gamma_{tot}$ the known $\chi_{c1}(1P)$ total width.

For the tensor charmonium state $\chi_{c2}(1P)$ the two-photon decay width is known. Assuming maximum helicity-2 contribution, corresponding with $r^{(2)} = 1$ and $r^{(0)} = 0$, this determines the helicity-2 coupling. We note that $r^{(0)} = 0$ matches the quark model result, see Eq.~(\ref{T:ratios}). In order to allow for a possible (small) non-zero value in experiment, we can approximate the product $r^{(0)} R_{{\cal T} }^{L}(0)$ appearing in SR$_2$ and SR$_3$ for small $r^{(0)} \ll r^{(2)}$ as (see Eq.~(A11) of Ref.~\cite{Danilkin:2016hnh}):
\begin{eqnarray}
r^{(0)} R_{{\cal T} }^{L}(0) &=& \frac{2 \sqrt{2}}{\sqrt{3}} \sqrt{r^{(0)}}  
\left( \frac{F_{{\cal T} \gamma ^*\gamma ^*}^{(0,L)}\left(0,0\right) }{F_{{\cal T} \gamma ^*\gamma ^*}^{(2)}\left(0,0\right) } \right),  \\
&\simeq & - \frac{2 \sqrt{2}}{\sqrt{3}} \sqrt{r^{(0)}}\,,\nonumber
\end{eqnarray} 
where in the last equality, the quark model ratio of Eq.(\ref{T:ratios}) has been used for the longitudinal amplitude. Furthermore, for the helicity-1 $\gamma^\ast \gamma$ coupling to a tensor charmonium state, we also adopt the quark model ratio from 
Eq.~(\ref{T:ratios}), i.e. 
\begin{eqnarray}
\left[ F^{(1)}_{{\cal T} \gamma^\ast \gamma^\ast}(0, 0)  \right]^2 = \frac{5 \, \Gamma _{\gamma \gamma }({\cal T})}{\pi\alpha ^2\,m_{\cal T}}.
\end{eqnarray}
This yields the relations:
\begin{eqnarray}
\mathrm{SR}_{2} \, [\mathrm{nb / GeV}^2] &\simeq& 0.10 \, 
- 0.15 \times  \left( 10^3 \, \tilde \Gamma_{\gamma \gamma} / \Gamma_{tot} \right)
\nonumber\\&& 
+0.14 \left [ 1 - 1.63 \sqrt{r^{(0)}} 
+ 0.5 \right]  
\nonumber\\&&+ 
I_{cont} (\mathrm{SR}_2),
\label{sr2num} \\\nonumber \\
\mathrm{SR}_{3} \, [\mathrm{nb}] &\simeq&
- 2.41  \, 
+ 1.79 \times  \left( 10^3 \, \tilde \Gamma_{\gamma \gamma} / \Gamma_{tot} \right)
\nonumber\\ && 
+1.81 \left[ - 1.63 \, \sqrt{r^{(0)}}   \, - 0.5 \right]
\nonumber\\&& 
+I_{cont} (\mathrm{SR}_3) ,
\label{sr3num}
\end{eqnarray}
where the terms on the {\it rhs} of SR$_2$ correspond with the contributions from $\chi_{c0}(1P)$, 
 $\chi_{c1}(1P)$, $\chi_{c2}(1P) [\Lambda = 2, 1, (0,L)]$, and continuum respectively, 
 and where the terms on the {\it rhs} of SR$_3$ correspond with the contributions from $\chi_{c0}(1P)$, 
 $\chi_{c1}(1P)$, $\chi_{c2}(1P) [\Lambda = 1, (0,L)]$, and continuum respectively. 
We will determine the unknown equivalent decay width $\tilde \Gamma_{\gamma \gamma}$ 
for the $\chi_{c1}(1P)$ state as well as a possible small non-zero $r^{(0)}$ value for the tensor charmonium state 
by the requirement that both sum rules are satisfied simultaneously by the three lowest charmonium bound states and by the continuum contributions.

\begin{table*}[t]
{\centering \begin{tabular}{|c|c|c|c|c|}
\hline
&&&& \\
& $m_M$  & $\Gamma_{\gamma \gamma} $  
&   
$\int ds \; \left( \frac{1}{s^2} \sigma_\parallel  + \frac{1}{s} \frac{\tau^a_{TL}}{Q_1 Q_2} \right)_{Q_i^2  = 0}$ 
&   $\int ds \; \left(  \frac{\tau_{TL}}{Q_1 Q_2} \right)_{Q_i^2  = 0}$   \\
&  [MeV] &  [keV] &   [nb / GeV$^2$]  &   [nb]   \\
\hline 
\hline
\quad $\chi_{c0}(1P)$  \quad   & \quad $3414.75 \pm 0.31$  \quad  & \quad  $2.34 \pm 0.19$ \quad   &   \quad $ 0.10 \pm  0.01$  \quad 
& \quad $ - 2.41 \pm 0.20 $
\\
\hline
\quad $\chi_{c1}(1P)$  \quad   & \quad $3510.66 \pm 0.07$ \quad   & \quad  -  \quad  &   $ - 0.15  \times  \left(  10^3 \, \tilde \Gamma_{\gamma \gamma} / \Gamma_{tot} \right) $  
& \quad $1.79 \times  \left( 10^3 \, \tilde \Gamma_{\gamma \gamma} / \Gamma_{tot} \right) $ \quad 
\\
\hline
\quad $\chi_{c2}(1P)$  \quad   & \quad $3556.20 \pm 0.09 $ \quad   & \quad  $0.53 \pm 0.04$  \quad  &   
& 
  \\
\quad   \quad $\Lambda=2$  & & & \quad $ 0.14 \pm 0.01 $ \quad & 0 \\
\quad   \quad $\Lambda=(0,T)$  & & & \quad $ \approx 0 $ \quad   & 0  \\
\quad   \quad $\Lambda=(0,L)$  & & & \quad $ (- 0.23 \pm0.02)\sqrt{r^{(0)}}$ \quad   &  $ (- 2.96\pm0.22) \sqrt{r^{(0)}} $ \\
\quad   \quad $\Lambda=1$  & & & \quad $ 0.07 \pm 0.01 $ \quad  & \quad $-0.91 \pm 0.07 $ \quad \\
\hline
\hline
continuum  contr.  &  & &  
\quad $ -0.040 \pm 0.014 $\quad & $0$ \\
\hline
\end{tabular}\par}
\caption{Contributions of the lowest $c \bar c$ mesons to the second sum rule (SR$_2$, 4th column) and third sum rule (SR$_3$, 5th column) of Eq.~(\ref{eq:sumrules}). We show the bound state contributions  based on the 2016 PDG values~~\cite{Olive:2016xmw} of the meson masses $m_M$ and their $2 \gamma$ decay widths $\Gamma_{\gamma \gamma}$. 
In the last line, we also show the duality estimate 
for the continuum contribution above $D \bar D$ threshold. 
}
\label{table_sr2sr3}
\end{table*}

To estimate the continuum contributions to the sum rules of Eqs.~(\ref{sr2meson0}, \ref{sr3meson0}), we will again use a duality argument by
replacing the integral for the $\gamma \gamma  \to X$ process (with $X$ any hadronic final state containing charm quarks)
by the corresponding integral for the perturbative $\gamma \gamma \to c \bar c$ process: 
\begin{eqnarray}
I_{cont} (\mathrm{SR}_2) & \approx & 
\int \limits_{s_D}^{\infty} \, ds \, 
\left( \frac{\sigma_\parallel }{s^2}  + \frac{1}{s} \frac{\tau^a_{TL}}{Q_1 Q_2} \right)_{Q_i^2  = 0} 
(\gamma \gamma \to c \bar c), 
\label{eq:duality2a} \nonumber\\
I_{cont} (\mathrm{SR}_3) & \approx & 
\int \limits_{s_D}^{\infty} \, ds \, 
\left(  \frac{\tau_{TL}}{Q_1 Q_2} \right)_{Q_i^2  = 0} 
(\gamma \gamma \to c \bar c).
\label{eq:duality3a}
\end{eqnarray}

The perturbative cross sections for the free $\gamma \gamma  \to c \bar c$ process were 
calculated in Ref.~\cite{Pascalutsa:2012pr} and were verified to satisfy SR$_2$ and SR$_3$ exactly. 
For SR$_2$ this implies 
\begin{eqnarray}
&&I_{cont} (\mathrm{SR}_2) 
\approx 
 - 4 \pi \alpha^2  e_c^4 \, N_c   \int \limits_{4 m_c^2}^{s_D} \frac{ds}{s^3} 
\left\{ - \left( 5 + \frac{6 m_c^2}{s} \right) \beta(s) \right. 
\nonumber\\ &&
\left. \hspace{0.5cm}+ 2 \left( 1 +  \frac{12 m_c^2}{s} - \frac{12 m_c^4}{s^2} \right) \, \ln \left( \frac{\sqrt{s}}{2 m_c} \left( 1 + \beta(s) \right)\right) 
\right\}\,.
\nonumber \\
&&
\label{eq:duality2b}
\end{eqnarray}
Using the physical value of the $D \bar D$ threshold, $s_D \approx 14$~GeV$^2$, 
and the PDG value $m_c = 1.27 \pm 0.03 $~GeV \cite{Olive:2016xmw}, 
we obtain: $I_{cont} (\mathrm{SR}_2) \approx - 0.040 \pm 0.004 $~nb/GeV$^2$, where the error results from the uncertainty range in the PDG value for $m_c$. However, from the analysis of the SRI, we know that the perturbative cross section contribution to the continuum would need to be changed by around $35\%$ to saturate the sum rule exactly. Therefore, for SRII we increase the error bar of the continuum contribution by around $35\%$ of its central value leading to $I_{cont} (\mathrm{SR}_2) \approx - 0.040 \pm 0.014 $~nb/GeV$^2$. For SR$_3$, the perturbative cross sections are zero when both photons are quasi-real, i.e. ~\cite{Pascalutsa:2012pr}:
\begin{eqnarray}
\left( \frac{\tau_{TL}}{Q_1 Q_2} \right)_{Q_i^2  = 0} 
(\gamma \gamma \to c \bar c) = 0.
\end{eqnarray}
Consequently, when using the free $\gamma \gamma  \to c \bar c$ process to estimate the continuum contribution, we obtain~:
$I_{cont} (\mathrm{SR}_3) = 0$.

When using the plane wave continuum contribution, the requirement that both sum rules of Eqs.~(\ref{sr2num}, \ref{sr3num}) are satisfied then yields for $\chi_{c1}(1P): \,{\tilde \Gamma_{\gamma \gamma}}/ \Gamma_{tot} \simeq \left(1.88 \pm 0.12 \right) \times 10^{-3} ,$ and for
$\chi_{c2}(1P): \sqrt{r^{(0)}}  \simeq 0.02 \pm 0.04,$
confirming that the $\Lambda = 0$ two-photon coupling for the ${\chi_{c2}(1P)}$ state is very small. We also made an estimate of the change due to the contribution of the states $\chi_{c0}(2P)$ and $\chi_{c2}(2P)$ states. We  found that their contribution leads to increase of $ \tilde \Gamma_{\gamma \gamma}/ \Gamma_{tot} $ for $\chi_{c1}(1P)$ between 12\% (based on the analysis of $SR_3$) and 20\% (based on the analysis of $SR_2$). We conservatively use the larger value as our error estimate, which leads to:
\begin{eqnarray}
&&  {\chi_{c1}(1P)}: \quad 
 \frac{\tilde \Gamma_{\gamma \gamma}}{ \Gamma_{tot}} \simeq \left(1.9 \pm 0.4 \right) \times 10^{-3}\,.
\label{eq:chic1_gaga}
 \end{eqnarray}
A more precise evaluation of these higher contributions will require $\gamma^*\gamma$ data above $D\bar{D}$ threshold.

Using the PDG 2016 value~\cite{Olive:2016xmw} for the total width 
$\Gamma \left[ {\chi_{c1}(1P)} \right] =   0.84 \pm 0.04~\mathrm{MeV}$, Eq.~(\ref{eq:chic1_gaga}) then yields:
\begin{eqnarray}
{\chi_{c1}(1P)}: \quad \tilde \Gamma_{\gamma \gamma}  \simeq  \left( 1.6 \pm 0.3 \right) \; \mathrm{keV}.
\label{eq:chic1_gaga2}
\end{eqnarray}

We can contrast our result for $\tilde\Gamma_{\gamma \gamma}$ of $\chi_{c1}(1P)$ with the quark model result for the charmonium states. At leading order, the non relativistic quark model prediction of  Eq.~(\ref{Gammas_QuarkModel}) leads to the two photon decay rates
\begin{eqnarray}\label{QM_1}
\tilde \Gamma_{\gamma \gamma}  \left[ {\chi_{c1}(1P)} \right]  &\simeq& \frac{5}{6} \,  \Gamma_{\gamma \gamma}  \left[ {\chi_{c2}(1P)} \right], \\
 &\simeq& \frac{2}{9} \,  \Gamma_{\gamma \gamma}  \left[ {\chi_{c0}(1P)} \right]\,.\label{QM_2}
\end{eqnarray}
Radiative corrections and relativistic effects are however expected to somewhat change such estimates. As it is shown in \cite{Ebert:2003mu, Badalian:2008bi}, the ratio of two photon decay widths of scalar and tensor states increases by including radiative corrections. Relativistic effects, on the other hand, partly compensate that change. In combination, these effects can change the leading order quark model predictions by at most 50\% (see Table 2 of \cite{Ebert:2003mu}), which we include as the uncertainty of the $\chi_{c1}(1P)$ estimate. Using the empirical $\Gamma_{\gamma \gamma}$ values for  ${\chi_{c0}(1P)} $ and ${\chi_{c2}(1P)} $ from Table~\ref{table_sr2sr3}
then yields the quark model prediction (the average between Eqs. (\ref{QM_1}) and (\ref{QM_2}))~:
\begin{equation}
\tilde \Gamma_{\gamma \gamma}  \left[ {\chi_{c1}(1P)} \right] \big|_{\rm{quark \, model}}  \simeq  \left( 0.48 \pm 0.24  \right) \; \mathrm{keV}.
\end{equation} 
One notices that our sum rule extraction of Eq.~(\ref{eq:chic1_gaga2}) yields an equivalent two-photon width 
$\tilde \Gamma_{\gamma \gamma}$ for the $\chi_{c1}(1P)$ state, which is  at variance with the quark model estimate by around $3\,\sigma$. 
Such prediction can be tested by data from the $e^+ e^-$ collider experiments, in particular Belle and BESIII, where the $\chi_{c1}(1P)$ can be produced in $\gamma^\ast \gamma$ collisions.

\section{Conclusions}

In this work, we have applied three forward light-by-light scattering sum rules to charmonium states. We have shown that these sum rules imply a cancellation between charmonium bound state contributions, which we quantified using their $\gamma \gamma$ decay widths, and continuum contributions above $D \bar D$ threshold. For the latter, we have provided a duality estimate, by replacing the $\gamma\gamma\rightarrow X$ cross sections, with $X$ denoting the sum over all allowed final states  entering the continuum parts of the sum rule integrals, by the corresponding perturbative $\gamma\gamma\rightarrow Q \bar{Q}$ cross sections. For the $\sigma_2 - \sigma_0$ sum rule, we have shown that the continuum contribution compensates to around  75 \% the sum of the lowest lying $\eta_c(1S), \chi_{c0}(1P)$, and $\chi_{c2}(1P)$ bound state contributions. For the higher states, we have shown that the estimate for the $\eta_c(2S)$, $\chi_{c0}(2P)$, and $\chi_{c2}(2P)$ states is nearly compensated within the error bar by the duality estimate in the interval $[4m_D^2,\,s_{D_2}]$, providing further support of our procedure. We have applied two further sum rules, in which at least one photon is quasi-real, to the charmonium states. The latter sum rules imply contributions of scalar, axial-vector and tensor mesons. We have shown that these sum rules allow to predict the yet unmeasured $\gamma^\ast \gamma$ coupling, with one longitudinal and one transverse photon, of the $\chi_{c1}(1P)$ state as: $\tilde \Gamma_{\gamma \gamma}  \simeq  \left( 1.6 \pm 0.30 \right) \; \mathrm{keV}$, or equivalently 
$ \tilde \Gamma_{\gamma \gamma} /  \Gamma_{tot} \simeq \left(1.9 \pm 0.4 \right) \times 10^{-3}$.
This prediction at variance with the quark model estimate by around $3\,\sigma$, and can be tested at present high-luminosity $e^+ e^-$ colliders. In view of our analysis, indicating important $\gamma^\ast \gamma$ cross section contributions above $ D \bar D$ threshold, the measurement of these cross sections in the region where a large number of new, so-called $XYZ$, states have been found in recent years, several of which observed in $\gamma \gamma$ collisions, is very promising to shed further light on the nature of these states.

\section*{Acknowledgements}
This work was supported by the Deutsche Forschungsgemeinschaft (DFG) 
in part through the Collaborative Research Center [The Low-Energy Frontier of the Standard Model (SFB 1044)], and in part through the Cluster of Excellence [Precision Physics, Fundamental
Interactions and Structure of Matter (PRISMA)].

\appendix 

\section{$\gamma^* \gamma^* \to \, ^1S_0, \, ^3P_0, \, ^3P_1, \, ^3P_2$ matrix elements in the quark model}
\label{Appendix_Quark_Model}

\begin{figure}
\centering
\includegraphics[width=0.50\textwidth]{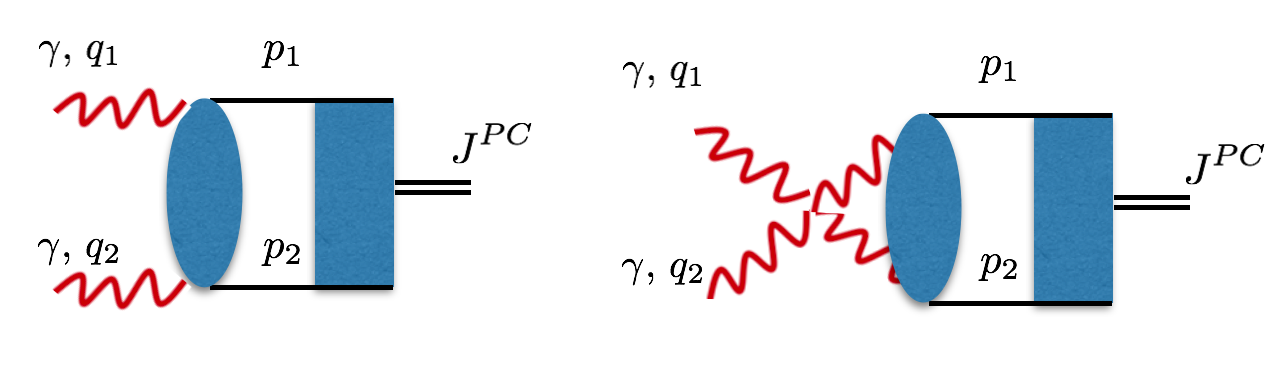}
\caption{ Diagrams illustrating the $\gamma^\ast \gamma^\ast \to Q \bar Q$ production mechanisms. 
\label{fig:quarkmodel}}
\end{figure}

In this appendix, we calculate the lowest lying $\gamma^* \gamma^* \to \, ^1S_0, \, ^3P_0, \, ^3P_1, \, ^3P_2$ production  matrix elements in the quark model. The transition matrix element in the quark model \cite{Ackleh:1991dy,Ebert:2003mu,Hanhart:2007wa} can be written as a convolution integral between the quarkonium wave function (w.f.) $\psi_{nl m_l}(\textbf{p})=\tilde{R}_{n\,l}(p)\,Y_{l\,m_l}(\theta,\phi)$ and the plane wave $\gamma^*\gamma^*\rightarrow Q\bar{Q}$ scattering amplitude $\langle \textbf{p} | M_{\lambda_1\,\lambda_2} | \textbf{q}\rangle$ as:
\begin{eqnarray}\label{RescattDiscr}
\langle J^{PC} | M_{\lambda_1\,\lambda_2} | \textbf{q}\rangle &&=\int \frac{d^3\,\textbf{p}}{(2\pi)^3}\sum_{s_1,s_2}\,\,\langle \textbf{p} | M_{\lambda_1\,\lambda_2}(s_1,s_2) | \textbf{q}\rangle\,\nonumber\\
&& \times \sum_{m_l, m_s} \langle J\,m_J| \,l\,m_l\,s\,m_s \rangle \\
&& \times \langle s\,m_s| 1/2\,s_1\,1/2\,s_2 \rangle\,\psi^*_{nlm_l}(\textbf{p})\,\frac{\sqrt{2 M}}{2E(\textbf{p})},\nonumber
\end{eqnarray}
where $m_Q$ is the mass of the quark, $E(\textbf{p})=\sqrt{\textbf{p}^2+m_Q^2}$, $M$ is the $Q\bar{Q}$ mass and $\lambda_1(\lambda_2)$ are the photon helicities. The $\gamma^*\gamma^*\rightarrow Q\bar{Q}$ amplitude to lowest order in $\alpha$ is given by
\begin{eqnarray}\label{AmplSpinorQED_0}
\langle\textbf{p} |M_{\lambda _1 \lambda _2}(s_1,s_2)|\textbf{q}\rangle&&=-i\,\sqrt{N_c}\,e_Q^2 \,e^2\epsilon _{\lambda_1\,\mu }\,\epsilon _{\lambda_2\,\nu }\,\bar{u}(p_1,s_1)
\nonumber\\
&&\times \left\{
\frac{\left(2\,p_1^{\mu }- \gamma ^{\mu }\gamma \cdot {q}_1\right)\gamma ^{\nu}}{\left(p_1-q_1\right)^2-m_Q^2}\right.
\\
&&\left.
+
\frac{\left(2\,p_1^{\nu }-\gamma ^{\nu }\gamma \cdot {q}_2 \right)\gamma ^{\mu }}{\left(p_1-q_2\right)^2-m_Q^2}\right\}v(p_2,s_2)\,,\nonumber
\end{eqnarray}
where $q_{1,2}$ are the momenta of the incoming photons, and $p_{1,2}$ are the momenta of the outgoing quarks. The relative momenta of the initial and final particles are denoted by $\textbf{q}=\frac{1}{2}(\textbf{q}_1-\textbf{q}_2)$ and $\textbf{p}=\frac{1}{2}(\textbf{p}_1-\textbf{p}_2)$, respectively (see Fig. \ref{fig:quarkmodel}). To evaluate (\ref{RescattDiscr}) we use Dirac spinors which are defined as 
\begin{eqnarray}
u(\textbf{p}, s)&=&\sqrt{E(\textbf{p})+m_Q} \left(
\begin{array}{c}
1\\
\frac{\mathbf{\sigma}\, \textbf{p} }{E(\textbf{p})+m_Q}\\
\end{array}
\right)\,\chi^{(s)}\,,\quad
\nonumber\\
v(\textbf{p}, s)&=&\sqrt{E(\textbf{p})+m_Q} \left(\begin{array}{c}
\frac{\mathbf{\sigma}\,\textbf{p}}{E(\textbf{p})+m_Q}\\
1\\
\end{array} \right)\,\eta^{(-s)}\,,\nonumber
\end{eqnarray}
with two-component Pauli spinors
\begin{eqnarray}
\chi^{(1/2)}&=&\left(\begin{array}{c}
1\\
0\\
\end{array} \right)\,, \quad 
\chi^{(-1/2)}=\left(\begin{array}{c}
0\\
1\\
\end{array} \right)\,,\quad
\nonumber\\
\eta^{(-1/2)}&=&\left(\begin{array}{c}
0\\
1\\
\end{array} \right)\,, \quad 
\eta^{(1/2)}=\left(\begin{array}{c}
-1\\
0\\
\end{array} \right)\,.\nonumber
\end{eqnarray}
The sign convention in $\eta^{(1/2)}$ is chosen so that the spinors are charge conjugates of each other \cite{Gross2008}. Below we assume nonrelativistic kinematics, such as $E(\textbf{p})\approx m_Q$ and $s=M^2\approx(2m_Q)^2$.

\begin{table*}[t]
\renewcommand{\arraystretch}{3.0}
\fontsize{9}{11}
{\centering 
\begin{tabular}{@{\extracolsep{\fill}}ccccc@{}}
\hline
State & $M_{+-}$ &  $M_{++}$  &   $M_{0+}$  &  $M_{00}$   \\
\hline 
\hline
${}^1S_0$  & 0 & $\sqrt{\frac{2 X}{\pi  m_Q}}\frac{1}{\nu}\,R_{n S}(0)$ & 0 & 0 \\
\hline
${}^3P_0$  & 0 & $\sqrt{\frac{2}{\pi }}\frac{X+4 m_Q^2\nu }{m_Q^{3/2} \nu ^2}\,R_{n P}'(0)$ & 0 & $-4\sqrt{\frac{2 m_Q}{\pi }}\frac{Q_1 Q_2}{\nu ^2} R_{n P}'(0)$\\
${}^3P_1$  & 0 & $\sqrt{\frac{3}{\pi }}\frac{Q_1^2-Q_2^2}{m_Q^{3/2} \nu }R_{n P}'(0)$ & $2\sqrt{\frac{3}{\pi  m_Q}}\frac{Q_1 \left(\nu +Q_2^2\right)}{\nu ^2}R_{n P}'(0)$ & 0\\
${}^3P_2$  & \quad$4\sqrt{\frac{6 m_Q}{\pi }}\frac{1}{\nu}R_{n P}'(0)$ \quad & \quad$\frac{2 \left(\nu  \left(2 m_Q^2-\nu \right)+Q_1^2 Q_2^2\right)}{\sqrt{\pi } m_Q^{3/2} \nu ^2}R_{n P}'(0)$\quad & \quad$- 2\sqrt{\frac{3}{\pi  m_Q}}\frac{Q_1 \left(\nu - Q_2^2 \right)}{\nu ^2}R_{n P}'(0)$\quad &\quad $8 \sqrt{\frac{m_Q}{\pi}} \frac{Q_1 Q_2}{\nu ^2} R_{n P}'(0)$\quad\\
\hline
\hline
\end{tabular}\par}
\caption{Matrix elements of the lowest $Q \bar Q$ mesons. All the expressions must be multiplied by the common factor $i \sqrt{3}\,e_Q^2\,e^2$.\label{table_Matrix_elements}}
\end{table*}

The results for the finite photon virtualities $q_i^2=-Q_i^2$ are collected in Table \ref{table_Matrix_elements}. The crossing variable $\nu$ and the virtual photon flux factor factor $X$ are defined as
\begin{eqnarray}
\nu&=&q_1\cdot q_2=\frac{1}{2}(s+Q_1^2+Q_2^2)\,,\nonumber\\
 X&=&(q_1\cdot q_2)^2-q_1^2\,q_2^2=\nu^2-Q_1^2Q_2^2\,.
\end{eqnarray}
Note, that the obtained helicity amplitudes are consistent with Ref. \cite{Schuler:1997yw} where slightly different conventions were used. 

The transition matrix elements at $Q_1^2=Q_2^2=0$ can be related to the  two-photon decay width of quarkonia. For the low-lying states with quatum numbers $J^{PC}=0^{++},\,0^{-+},\,1^{++}$ and $2^{++}$ one obtains
\begin{eqnarray}\label{Gammas_QuarkModel}
&&\Gamma_{\gamma\gamma}({\cal P})=\frac{3\,\alpha^2\,e_Q^4}{m_Q^2}|R_{n S}(0)|^2\,,\nonumber\\
&&\Gamma_{\gamma\gamma}({\cal S})=\frac{27\,\alpha ^2\,e_Q^4}{m_Q^4}|R_{n P}'(0)|^2\,, \nonumber \\
&&\Gamma_{\gamma\gamma}({\cal T}(\Lambda =2))=\frac{36\,\alpha^2\,e_Q^4}{5\,m_Q^4}|R_{n P}'(0)|^2\,,
\nonumber \\
&&\tilde \Gamma_{ \gamma \gamma}({\cal A}) \equiv \lim \limits_{Q_1^2 \to 0} \, \frac{m_{\cal A}^2}{Q_1^2} \, \frac{1}{2} \,
\Gamma \left( {\cal A} \to \gamma^\ast_L \gamma_T \right)\nonumber \\
&&\qquad\qquad =\frac{6\,\alpha^2\,e_Q^4}{m_Q^4}|R_{n P}'(0)|^2\,,
\end{eqnarray}
where $R_{n l}(0)$ denotes the wave function at the origin and $R'_{n l}(0)$ the derivative of the wave function at the origin. Using the relations between the helicity amplitudes $M_{\lambda_1,\lambda_2}$ and the transition form factors given in Appendix C of Ref. \cite{Pascalutsa:2012pr} one can derive the set of ratios in the quark model. For the scalar mesons one obtains
\begin{eqnarray}\label{S:ratios}
R_{{\cal S} }^{L}(0)\equiv \frac{F_{{\cal S} \gamma ^\ast\gamma ^\ast}^L(0,0)}{F_{{\cal S} \gamma ^\ast\gamma ^\ast}^T(0,0)}=\frac{2}{3}\,,
\end{eqnarray}
while for the tensor mesons the following ratios are useful 
\begin{eqnarray}\label{T:ratios}
&&\frac{F_{{\cal T} \gamma ^\ast\gamma ^\ast}^{(0,L)}(0,0)}{F_{{\cal T} \gamma ^\ast\gamma ^\ast}^{(2)}(0,0)}=-1,\nonumber\\
&&r^{(0)}=\frac{\Gamma_{\gamma\gamma}({\cal T}(\Lambda=0))}{\Gamma_{\gamma\gamma}({\cal T})}=0\,,  \\
&& \left[ F_{{\cal T} \gamma ^*\gamma ^*}^{(1)}\left(0,0\right) \right]^2 =  \left[ \frac{M_{0+}}{e^2\,Q_1\frac{1}{\sqrt{2}}\left(\frac{2X}{\nu \,m_{{\cal T}}^2}\right)} \right]^2_{Q_i^2=0}\nonumber\\
&&\hspace{2.5cm}=\frac{5\,\Gamma_{\gamma\gamma}({\cal T})}{\pi\,\alpha^2\,m_{{\cal T}}}\,,\nonumber
\end{eqnarray}
where $\Gamma_{\gamma\gamma}({\cal T})=\Gamma_{\gamma\gamma}({\cal T}(\Lambda=0))+\Gamma_{\gamma\gamma}({\cal T}(\Lambda=2))$.

\end{document}